\documentclass[letterpaper, 10 pt, conference]{ieeeconf}  

\IEEEoverridecommandlockouts                              
\pdfoutput=1

\usepackage{graphicx} 
\usepackage{amsmath} 
\usepackage{amssymb}  
\usepackage[usenames, dvipsnames]{color}
\usepackage{url}
\usepackage{siunitx}
\usepackage{lipsum}
\usepackage{booktabs}
\usepackage{multirow}
\usepackage{subcaption}
\usepackage{float}
\usepackage{cite}

\usepackage{caption}
\newcommand{\numberOfTestSubjects}{\textcolor{black}{4} }

\title{\LARGE \bf
Force-Ultrasound Fusion: \\
Bringing Spine Robotic-US to the Next ``Level"
}

\author{Maria Tirindelli$^{*,1}$, Maria Victorova$^{*,2}$, Javier Esteban$^{*,1}$, Seong Tae Kim$^{*,1}$,\\
David Navarro-Alarcon$^{2}$, 
Yong Ping Zheng$^{2}$ and Nassir Navab$^{1,3}$
\thanks{$^{*}$These authors contributed equally to this work.}
\thanks{$^{1}$Computer Aided Medical Procedures, Technische Universit\"at M\"unchen, Munich, Germany {\tt\small maria.tirindelli@tum.de}}
\thanks{$^{2}$ The Hong Kong Polytechnic University, Hung Hom, Hong Kong}
\thanks{$^{3}$Computer Aided Medical Procedures, Johns Hopkins University, Bal- timore, MD, USA}
\thanks{This research was funded by the Bayerische Forschungsstiftung,Grant DOK-180-19.
We thank ImFusion GmbH (Munich, Germany) for their image processing framework and continuous support.}
}

\begin{document}

\maketitle
\thispagestyle{empty}
\pagestyle{plain}

\begin{abstract}

Spine injections are commonly performed in several clinical procedures. The localization of the target vertebral level (i.e. the position of a vertebra in a spine) is typically done by back palpation or under X-ray guidance, yielding either higher chances of procedure failure or exposure to ionizing radiation. Preliminary studies have been conducted in the literature, suggesting that ultrasound imaging may be a precise and safe alternative to X-ray for spine level detection. However, ultrasound data are noisy and complicated  to  interpret. In this study, a robotic-ultrasound approach for automatic vertebral level detection is introduced. The method relies on the fusion of ultrasound and force data, thus providing both ``tactile" and visual feedback during the procedure, which results in higher performances in presence of data corruption.
A robotic arm automatically scans the volunteer's back along the spine by using force-ultrasound data to locate vertebral levels.
The occurrences of vertebral levels are visible on the force trace as peaks, which are enhanced by properly controlling the force applied by the robot on the patient back. Ultrasound data are processed with a Deep Learning method to extract a 1D signal modelling the probabilities of having a vertebra at each location along the spine. Processed force and ultrasound data are fused using a 1D Convolutional Network to compute the location of the vertebral levels. The method is compared to pure image and pure force-based methods for vertebral level counting, showing improved performance. In particular, the fusion method is able to correctly classify 100\% of the vertebral levels in the test set, while pure image and pure force-based  method  could  only  classify  80\%  and  90\%  vertebrae, respectively. The potential of the proposed method is evaluated in an exemplary simulated clinical application.

\end{abstract}

\section{INTRODUCTION}\label{sec:introduction}
\begin{figure*}[h]

    \centering
    \begin{subfigure}[t]{0.54\linewidth}
        
        \includegraphics[width=\linewidth]{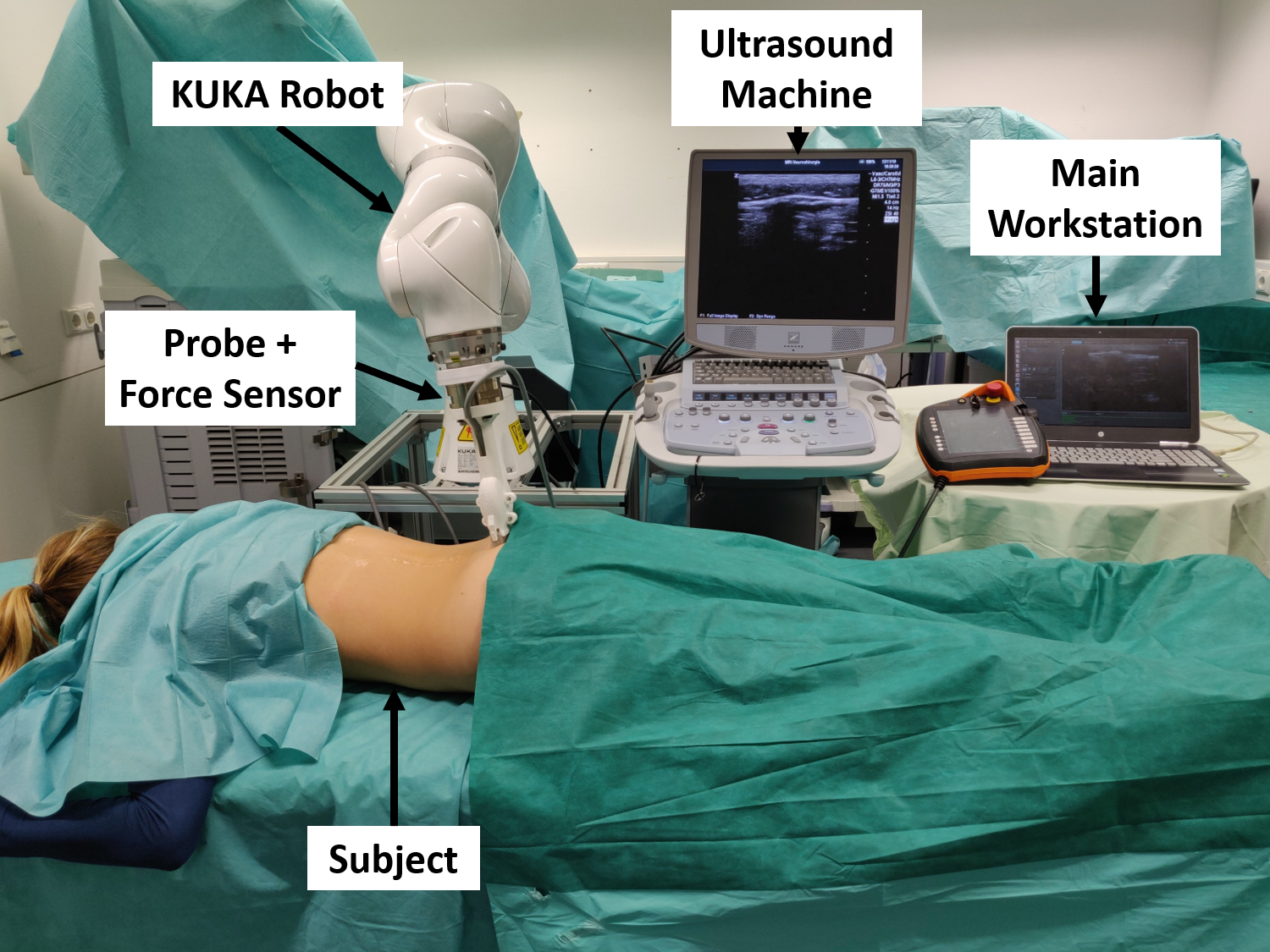}
        \caption{ }
        \label{fig:setup_system}
    \end{subfigure}\hspace{0.05\linewidth}%
    \begin{subfigure}[t]{0.3\linewidth}
        
        \includegraphics[width=\linewidth]{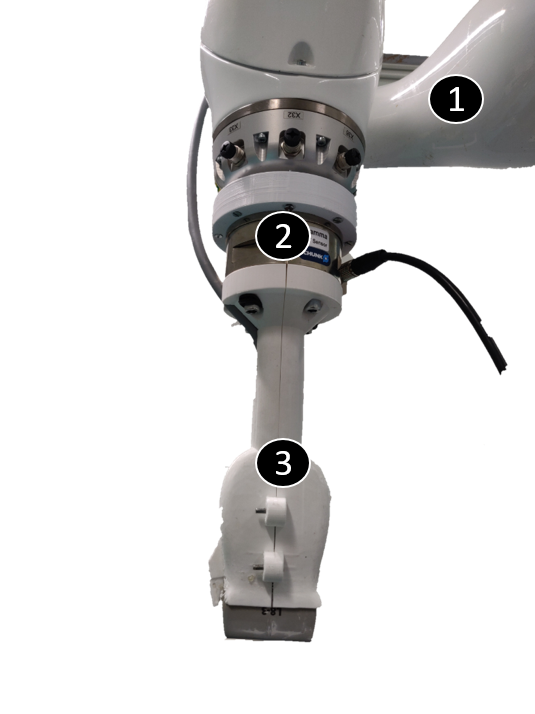}
        \caption{ }
        \label{fig:setup_probe}
    \end{subfigure}
    \caption{a) The Robotic Ultrasound System for Vertebral Level Classification Setup. b) The Robot end-effector configuration: 1 - Robotic arm, 2 - External force sensor, 3 - Ultrasound linear probe + 3D printed probe holder.}
    \label{fig:setup}
\end{figure*}

\newcolumntype{C}[1]{>{\centering\arraybackslash}m{#1}}
\setlength{\tabcolsep}{0.5em} 
{\renewcommand{\arraystretch}{1.5}

\begin{table*}[t]
\centering
\caption{\MakeUppercase{Dataset table with correspondent size, data and sensor settings}}
\begin{tabular}{ | C{1.5cm} | C{1.5cm}| C{2.5cm}| C{2.5cm}| C{2cm} | C{2cm}| C{2cm}|}
\hline
\textbf{Dataset} & \textbf{N. Subjects} & \textbf{Acquired Data} & \textbf{Probe Orientation} & \textbf{US Parameters} & \textbf{Applied Force [N]}  & \textbf{Robot Speed} [mm/s] \\
\hline
Dataset 1  & 19  & B-Mode Linear US &  Transverse & Gain = 92\% \newline Freq. = 14 Hz \newline Depth = 4cm & 2 & 20 \\
\hline
Dataset 2  & 13  & B-Mode Linear US \newline Force Data & Transverse & Gain = 92\% \newline Freq. = 14 Hz \newline Depth = 4cm & [2, 10, 15] & [12, 20, 40] \\
\hline
Dataset 3  & 19  & B-Mode Convex US &  Paramedian-Sagittal & Gain = 92\% \newline Freq. = 14 Hz \newline Depth = 7cm & 2 & 5\\
\hline
\end{tabular}

\label{tab:datasettable}
\end{table*}

}

Lumbar spinal injections are commonly performed in different clinical procedures as facet joint or epidural injections \cite{alexander2019lumbosacral, skaribas2019lumbar}. Such procedures typically require the correct localization of the target vertebra to effectively release pharmaceuticals. In clinical practice, vertebral level detection is achieved either through palpation or X-ray guidance. Although X-ray guidance can improve the overall precision of the procedure, the use of ionizing radiation is considered a hazard for the patient and especially for the clinicians and assistants. On the other hand, the accuracy of the palpation technique is lower, especially for less experienced clinicians. Furthermore, the incorrect chosen level of injection can lead to avoidable complications, such as headaches, nerve damage, and paralysis \cite{Boon2004}.

Ultrasound (US)  has proven to be an alternative to X-ray, providing precise guidance and preventing patients and clinicians from unnecessary radiation \cite{Galiano317,Evansa2015,wu2016effectiveness}. Despite being real-time and non-invasive, ultrasound guidance is particularly challenging in spine procedures due to artifacts and noise caused by the curvature of the spinal bones and the layer of soft tissue covering the spine. To address these issues, various authors have proposed to use image processing techniques to support the clinician in the detection of vertebral levels.

In \cite{Hetherington2017b} a method is proposed to automatically classify images acquired during manual ultrasound-guided epidural injections. In this work, a Convolutional Neural Network (CNN) is used to classify the acquired images as either ``vertebra" or ``intervertebral gap" and State Machine is implemented to refine the results.
In \cite{Kerby2008} and \cite{Yu2015} panorama image stitching is used to obtain a 2-Dimentional (2D) representation of  vertebral laminas along the spine in the paramedian-sagittal plane. In \cite{Kerby2008}  a set of filters are applied to the panorama image to enhance bony structures. Local minimums in the resulting pattern are extracted and labelled as vertebrae.  In \cite{Yu2015} the identification of vertebrae is performed on the panorama image using a template matching approach. 

The aforementioned methods provide support tools for the interpretation of ultrasound data during manual injection procedures. However, they still rely on the operator's skills to manually find correspondence between ultrasound images and patient anatomy. Few studies have been conducted to evaluate the potential of robots integration in the clinical environment for injection procedures.
In \cite{Esteban2018}, a robotic-ultrasound system for precise needle placement is described in an initial clinical study. In this study, a robotic system with a calibrated ultrasound probe is used to scan the patient back. The acquired US volume is then used by the operator to select the needle insertion path. The manipulator, equipped with a calibrated needle holder, moves to the desired insertion point to offer visual guidance during the insertion.
Although showing promising results, these systems still rely on the operator in the interpretation of ultrasound images. Furthermore, they do not provide any tactile feedback, which, for the standard procedure, is given by palpation.

The contribution of this work is a robotic-ultrasound approach combining force and ultrasound data for automatic lumbar vertebral level classification in the spine. The target spinal region is the lumbar region (i.e. vertebrae levels from L5 to L1), where spinal injections commonly take place.  Force feedback reproduces the tactile information the operator can get through palpation while ultrasound images provide continuous visual feedback during the procedure.
Compared to the previously presented methods for vertebrae level classification, the proposed approach combines the benefits of both robotics and standard procedures. Furthermore, it does not only rely on visual feedback, but it exploits multiple sensors information.
It is demonstrated that fusing ultrasound and force data ensures higher performances of the method in the presence of data corruption and single-sensor misclassifications. The potential of the proposed approach is explored for an example application, i.e. automatic target plane detection for facet injection procedures.

\section{Methods} \label{sec:methods}



\subsection{Materials and Experimental Setup}
\label{sec: materials and experimental Setup}
The system consists of a main workstation (Intel i7, GeForce GTX 1050 Mobile), a robotic arm certified for human interaction (KUKA  LBR  iiwa  7  R800) combined with a Six-Axis Force/Torque Sensor System FTD-GAMMA (SCHUNK GmbH \& Co. KG) and a Zonare z.one ultra sp Convertible Ultrasound System with an L8-3 linear probe, with purely linear and steered trapezoidal imaging (Fig. \ref{fig:setup}).
The  ultrasound  system  is  connected  to the  main  workstation  through  an Epiphan DVI2USB 3.0 frame-grabber (Epiphan Systems Inc. Palo Alto, California, USA), with an 800x600 resolution and a sampling frequency of 30 fps.
Deep Learning models were trained on an NVIDIA Titan V 12 GB HBM2, using Pythorch 1.1.0 as Deep Learning framework for both training and inference. ImFusion Suite Version 2.9.4 (ImFusion GmbH, Munich, Germany) is used for basic image processing and visualization. 

Three different datasets were used for training of Deep Learning models and testing. The datasets were acquired for different subjects with different ultrasound, robot force and speed settings. The acquisition was performed in the lumbar region, from L5 to L1. The Body Mass Index (BMI) of the scanned subject is in the range 20-30 for all the 3 databases. The dataset size and acquisition parameters are reported for the three datasets in Table \ref{tab:datasettable}.

\begin{figure*}[]
    \centering
    \includegraphics[width=\linewidth]{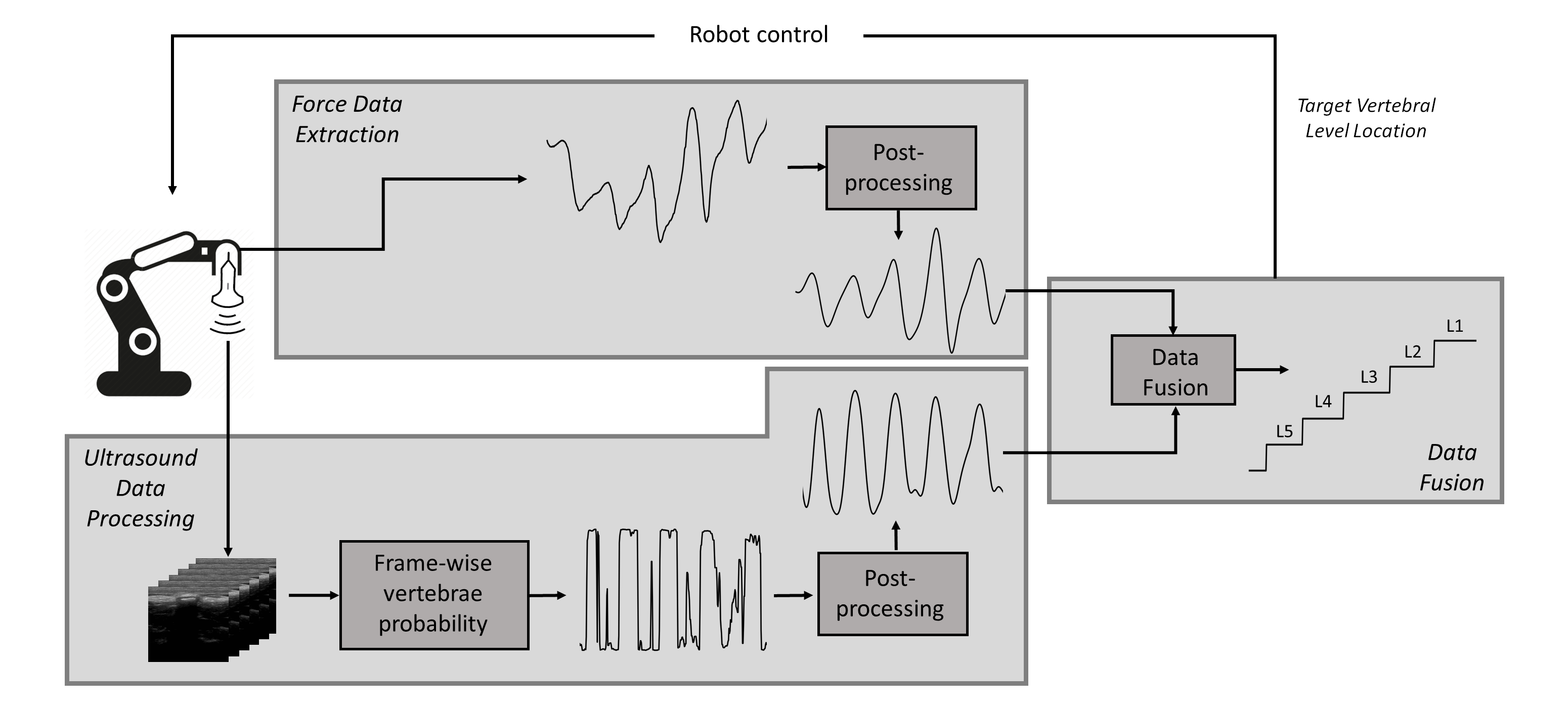}
    \caption{The method pipeline, including force data extraction, ultrasound data processing and the fusion method.}
    \label{fig:System_diagram}
\end{figure*}

In Fig. \ref{fig:System_diagram} a flowchart of the method is shown. In Sec. \ref{sec:scanningprocedure}, \ref{sec:forceextraction}, \ref{sec:imageprocessing} and \ref{sec:datafusion} a detailed description of each pipeline step is provided.

\subsection{Scanning Procedure}
\label{sec:scanningprocedure}

\begin{figure}[]

    \centering
    \begin{subfigure}[t]{0.454\linewidth}
        
        \includegraphics[width=\linewidth]{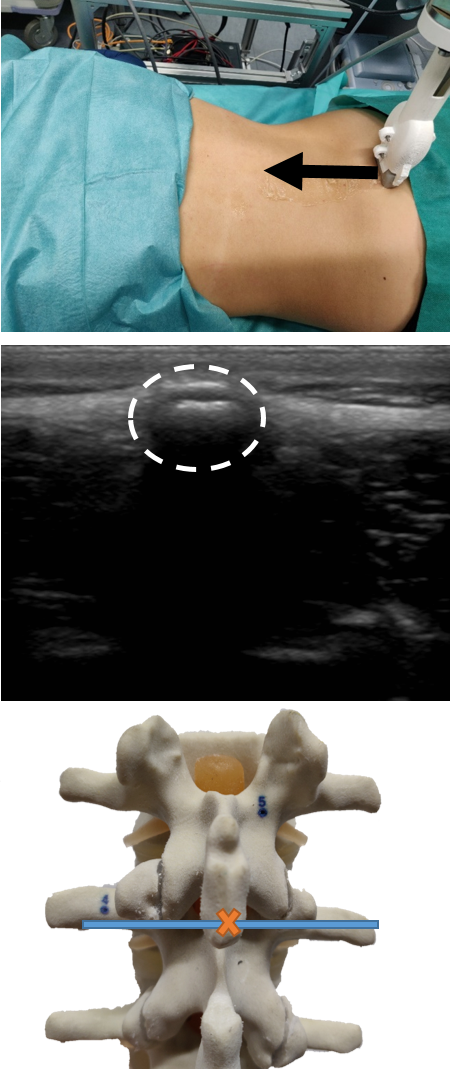}
        \caption{Data acquisition with a probe in transverse orientation and the respective ultrasound image of the spinous process.}
        \label{fig:scanningprocedure_subplot_spinous}
    \end{subfigure}\hspace{0.05\linewidth}%
    \begin{subfigure}[t]{0.45\linewidth}
        
        \includegraphics[width=\linewidth]{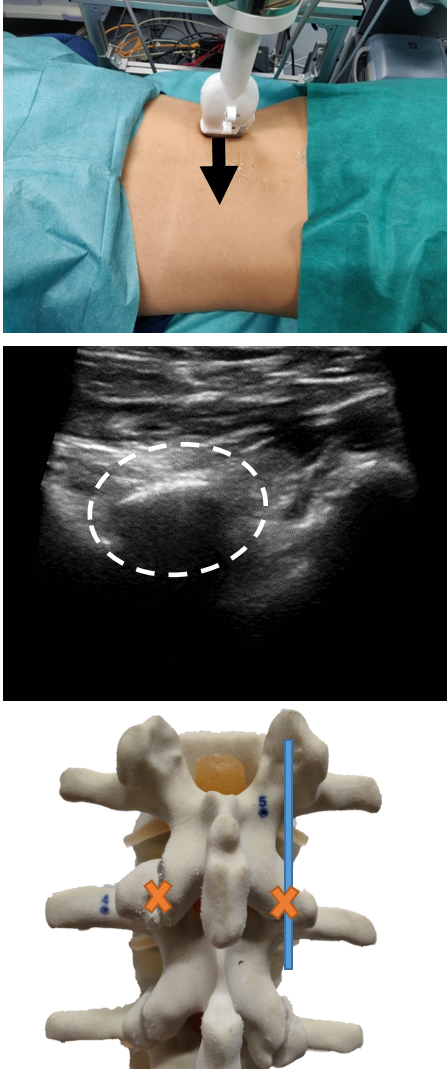}
        \caption{Data acquisition with the probe in paramedian sagittal orientation and the respective ultrasound image of the facet joint.}
        \label{fig:scanningprocedure_subplot_facet}
    \end{subfigure}
    \caption{Robot Trajectory during the procedure (arrows), target anatomies (dash line) and corresponding ultrasound images of acquired anatomies with planes of scanning (blue line).}
    \label{fig:scanningprocedure}
\end{figure}

Before starting the procedure, the  robotic  arm  is  manually  placed  at the level of the sacrum with a transverse probe orientation. After probe placement, the robot starts moving in an upward direction towards the subject's head, while force and ultrasound data are simultaneously collected (Fig. \ref{fig:scanningprocedure_subplot_spinous}). The subjects are asked to hold their breath for the whole duration of the scan (around 10 sec.), which is comparable to the breath-hold time of standard imaging procedures, as abdominal MRI or PET/CT \cite{van2012motion,pepin2014management}. 
Once the scan is completed, the collected data are processed, to provide the location of the vertebral level at which the injection must be performed. The robot is redirected to the target vertebra, where it can perform additional maneuvers depending on the clinical application. In the reported showcases (integration of the counting system for facet plane detection), the robot performs a further 90$^{\circ}$ rotation about the z-axis and acquires an ultrasound scan of the target vertebra with the probe in paramedian-sagittal orientation (Fig. \ref{fig:scanningprocedure_subplot_facet}).

 \begin{figure}[]
    \centering
    \includegraphics[width=\linewidth]{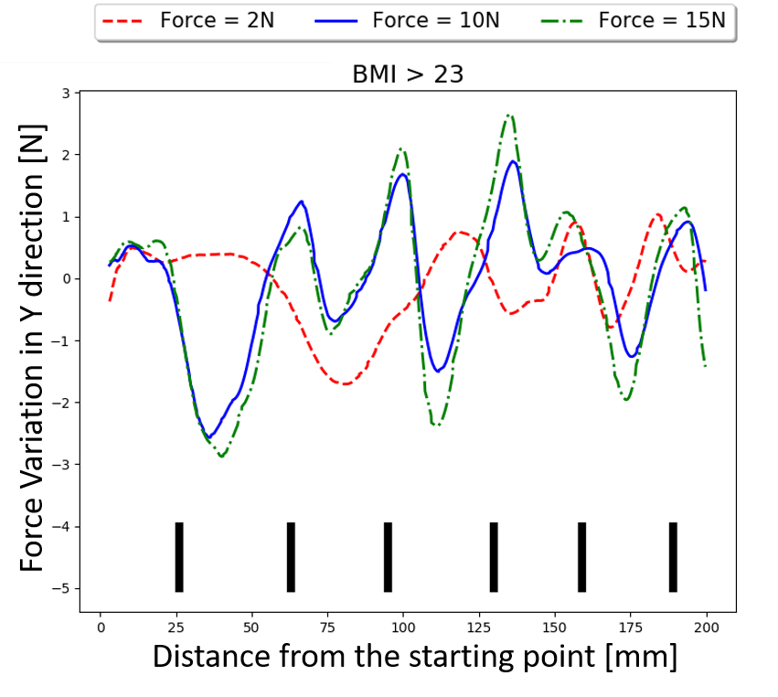}
    \caption{The force signal recorded in the y-axis with 3 different values (2, 10 and 15 N) of the z-force applied by the robot for subjects with ${BMI > 23}$.}
    \label{fig:forcecomparison_hightBMI}
\end{figure}

\begin{figure}[]
    \centering
    \includegraphics[width=\linewidth]{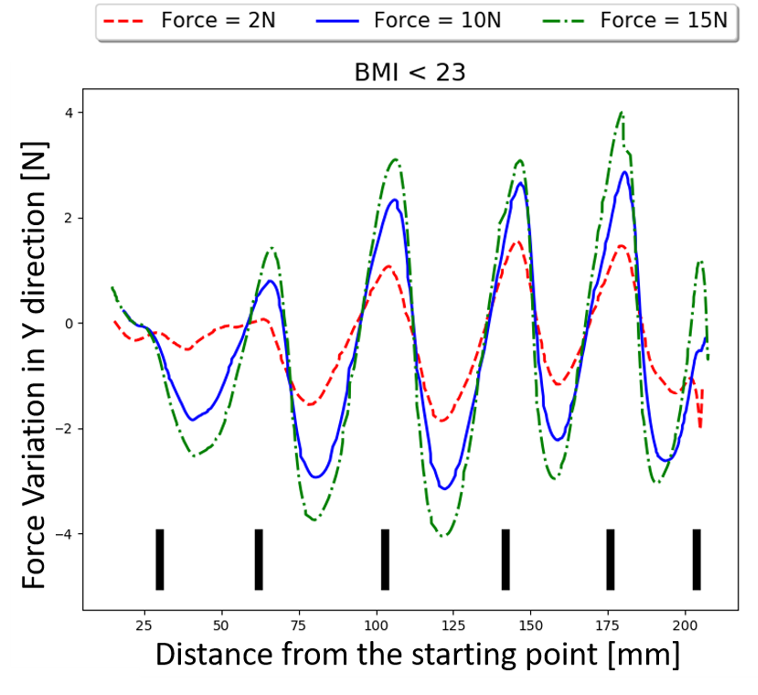}
    \caption{The force signal recorded in the y-axis with 3 different values (2, 10 and 15 N) of the z-force applied by the robot for subjects with ${BMI < 23}$.}
    \label{fig:forcecomparison_lowBMI}
\end{figure}

\begin{figure}
  \begin{subfigure}[b]{0.40\columnwidth}
    \includegraphics[width=\linewidth]{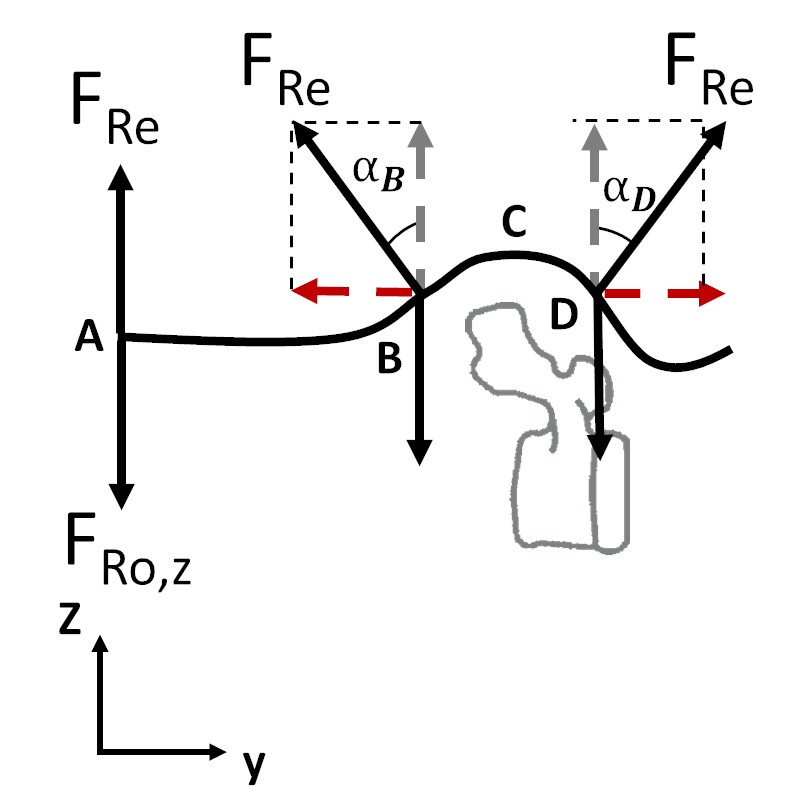}
    \caption{}
    \label{fig:forceinterraction}
  \end{subfigure}
  \begin{subfigure}[b]{0.59\columnwidth}
    \includegraphics[width=\linewidth]{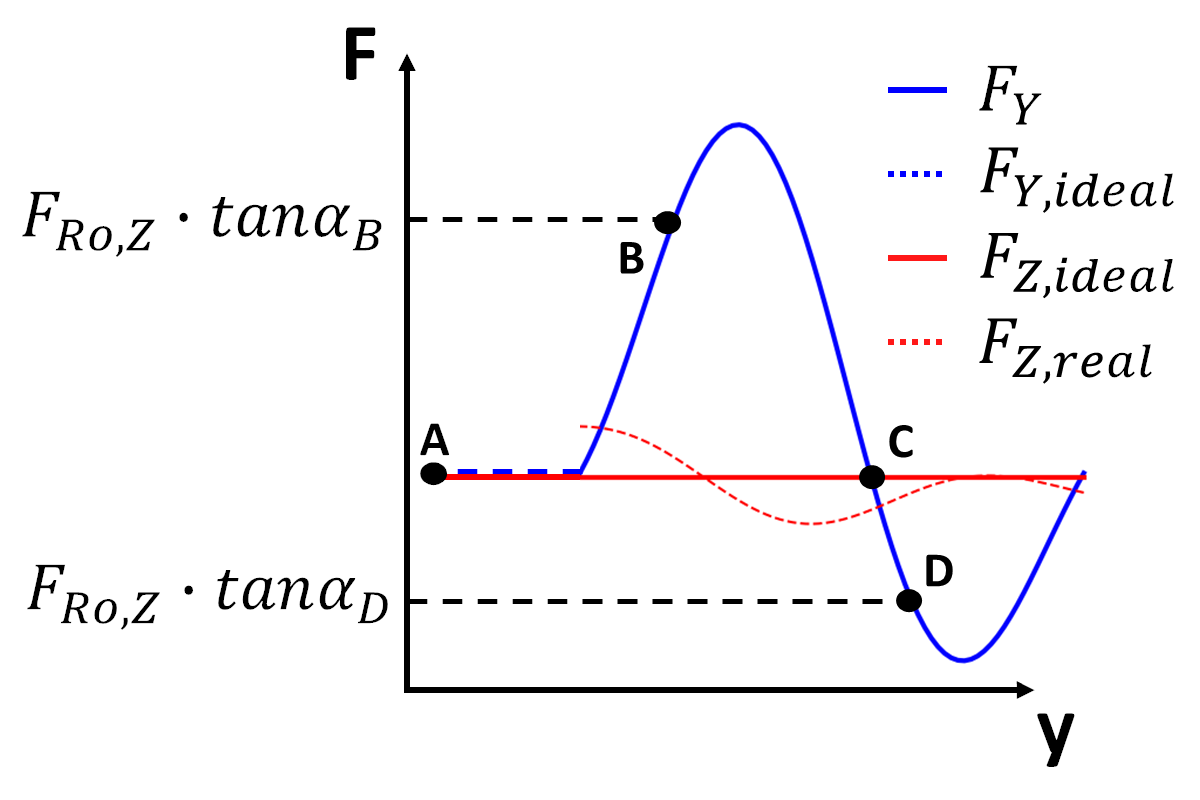}
    \caption{}
    \label{fig:forcetrace}
  \end{subfigure}
  \caption{(a) The modelled interaction between robot and patient back during the robotic scanning procedure. (b) Z component (red) and Y component (blue) of the force signal recorded over a single vertebra.}
  \label{forcefigure}
\end{figure}

\subsection{Force Data Extraction}
\label{sec:forceextraction}
In Fig. \ref{fig:forceinterraction}, a model of the vertebra-robot interaction is provided. 
In absence of vertebrae, the robot moves on a surface (the patient back) which can be considered flat. The reaction force is directed along the z-axis and its modulus balances the force applied by the robot, which is constant and set prior to the acquisition (Point A). In correspondence to a vertebra, the local direction of the subject back changes yielding to the generation of a non-null y-axis component of the reaction force (point B).
Once the vertebral peak has been reached (point C), the inclination of the plane changes again (point D) leading to the generation of non-null y-component of the reaction force, with an opposite sign with respect to point B. When the original surface direction is recovered, the y-component of the reaction force vanishes and the initial force value is recovered. 
The variations in the force y-component due to reaction forces are recorded by the force sensor and result in a very characteristic pattern in the force trace (Fig. \ref{fig:forcetrace}). This pattern can be used to count the vertebral levels while the patient back is scanned. In Fig. \ref{fig:forcetrace}, a plot of the y-component of the force signal is provided, in relation to the points A, B and C.

The recorded force in the y-direction ($F_y$) is pre-processed to remove the low-frequency drift, appearing due to the robot initial and final acceleration/deceleration. Drift removal is done by subtracting from the original signal its filtered version obtained applying a second-order Butterworth filter with cutoff frequency at 0.05 Hz. The ``un-drifted" signal is then low pass-filtered with a second-order Butterworth filter with cutoff frequency at 0.3 Hz, normalized between 0 and 1 and re-sampled in equally spaced space-grid.

As mentioned above, the force applied by the robot along the z-direction (${F_{Ro,z}}$) is constant and manually set before the acquisition takes place. The robot complies to the Force Control Scheme as described in \cite{Zettinig2017}. The value of the force z-component has a notable impact on the quality of the force signal recorded along the y-axis (${F_y}$) and on the visibility of vertebral patterns. 
In particular, higher values of ${F_{Ro,z}}$ lead to more visible and defined vertebral spikes. However, high values of ${F_{Ro,z}}$ also result in less comfort for the subjects, especially for those with a thin muscle/fat layer. 
In this study, the quality of the force signal recorded along the spine direction is evaluated for three different values of ${F_{Ro,z}}$ on a group of 13 subjects with BMI ranging from 20 to 30 (Dataset 2). The selected force values are comparable to those which are used in clinical experimentation \cite{Esteban2018}.  Each subject was asked to report the comfort level of the procedure on a scale ranging from 1 to 4, designed in the following way: 
1 - very uncomfortable,
2 - uncomfortable,
3 - slightly uncomfortable,
4 - comfortable.
For none of the subject, the procedure resulted to be ``very uncomfortable" or ``uncomfortable". However, subjects with lower BMI tended to rate the procedure performed with ${F_z = 15 N}$ as slightly uncomfortable. 
For this reason, the force applied by the robot along the z-axis is set to ${10 N}$ for subjects with lower BMI (${BMI < 23}$) and to ${15N}$ for subjects with higher BMI (${BMI > 23}$). 
In Fig. \ref{fig:forcecomparison_hightBMI} and Fig. \ref{fig:forcecomparison_lowBMI}, the force signal are reported for 3 different values of ${F_z}$ (i.e. ${2 N, 10 N, 15 N}$) for two subjects with different BMI. For both subjects, the amplitude of the spikes in the force trace increases with increasing force. However, for the subject with lower BMI, the spikes are still clearly recognizable in the signals obtained with lower pressures along the z-direction. 


\begin{figure*}[]

    \centering
    \begin{subfigure}[t]{0.31\linewidth}
        
        \includegraphics[width=\linewidth]{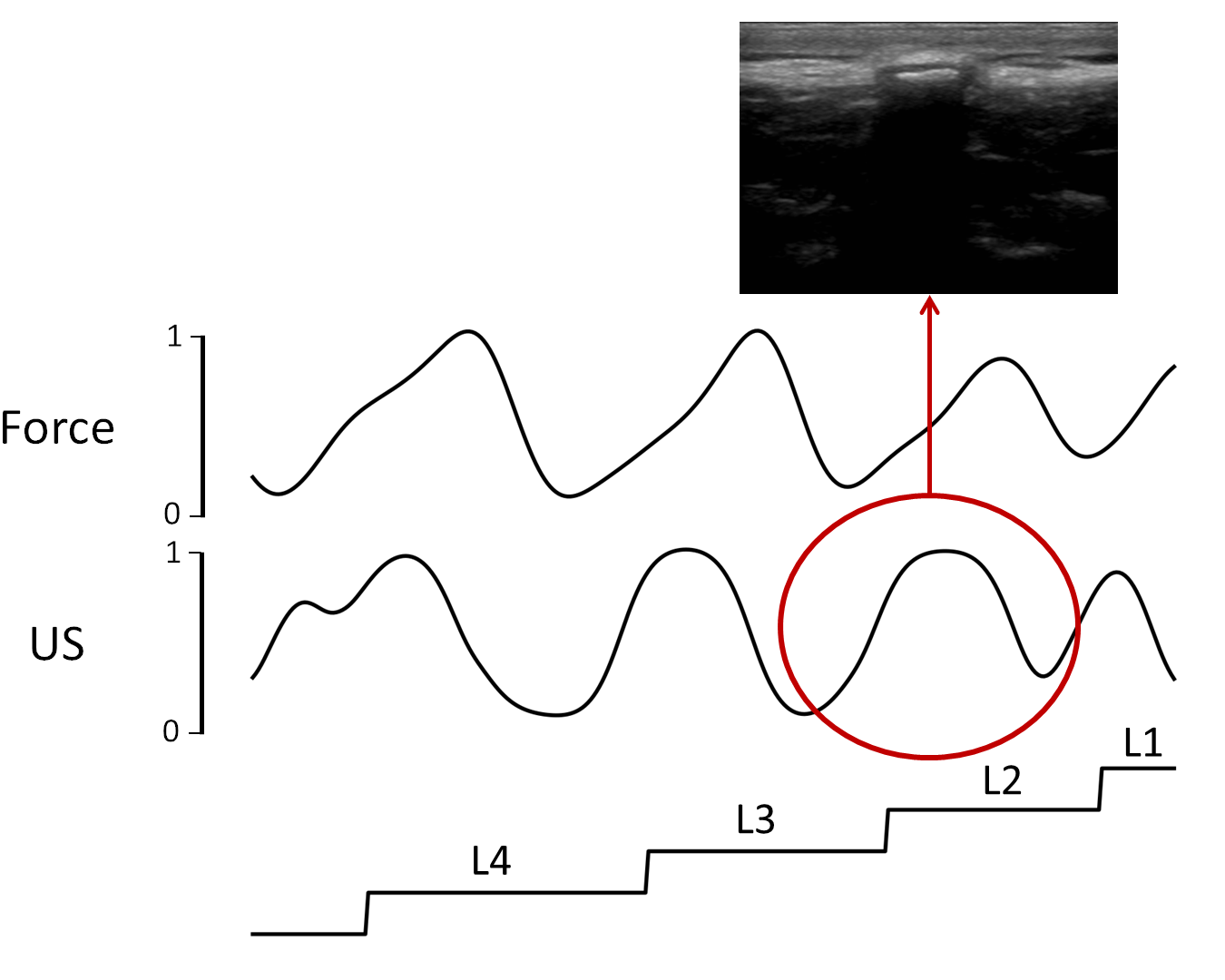}
        \caption{}
        \label{fig:goodforcegoodimg}
    \end{subfigure}%
    \begin{subfigure}[t]{0.29\linewidth}
        
        \includegraphics[width=\linewidth]{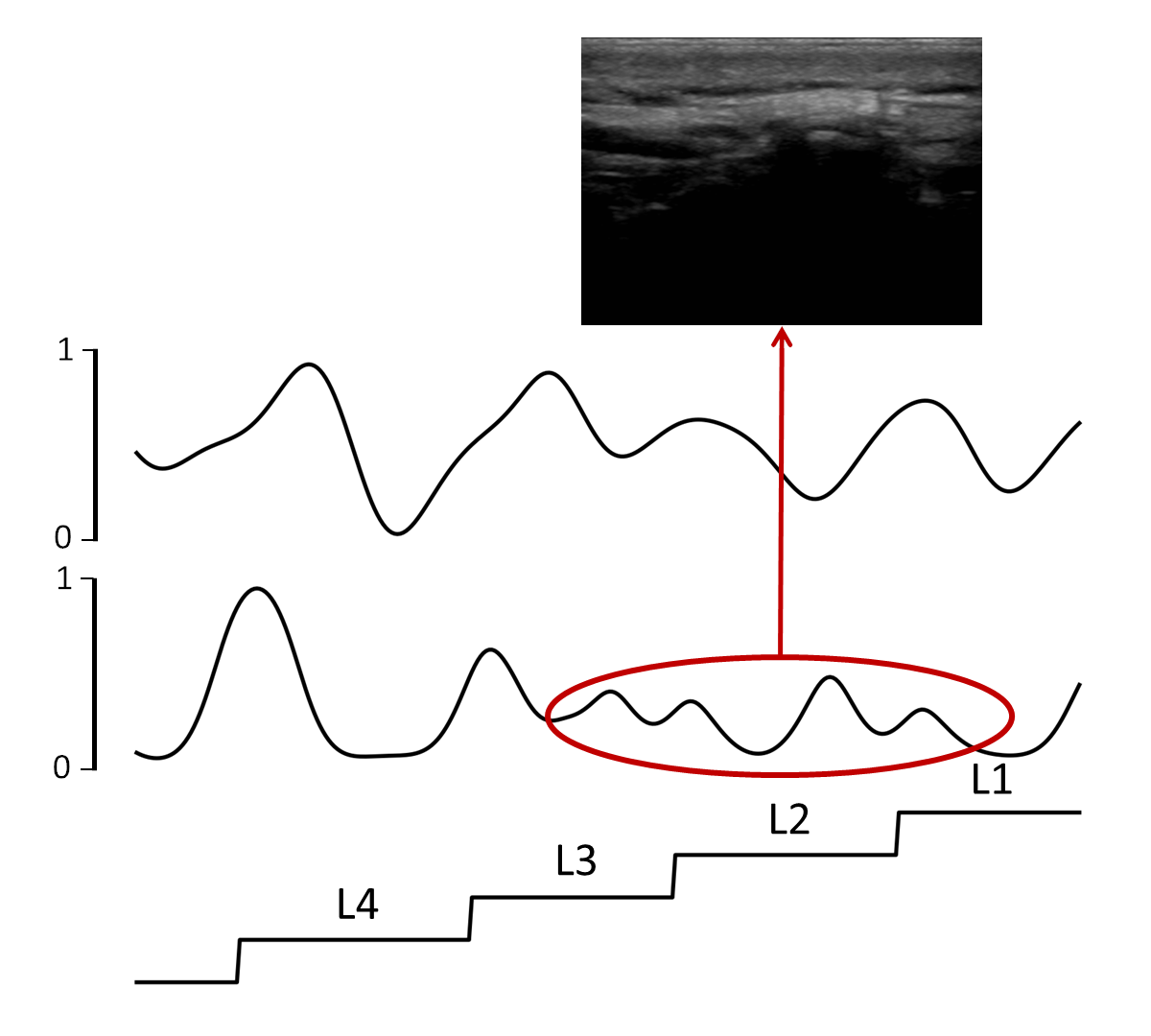}
        \caption{}
        \label{fig:goodforcebadimage}
    \end{subfigure}
    \begin{subfigure}[t]{0.29\linewidth}
        
        \includegraphics[width=\linewidth]{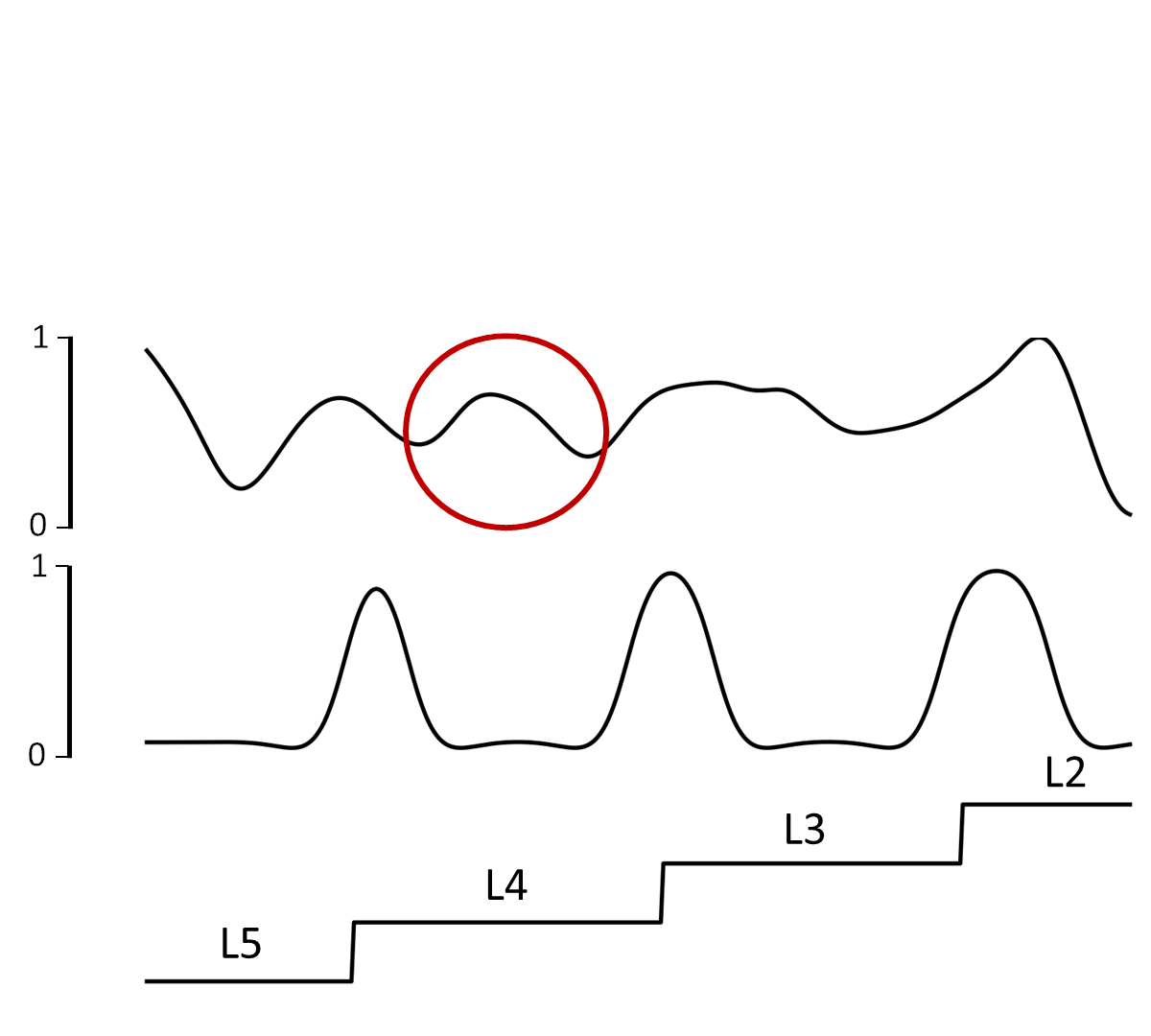}
        \caption{}
        \label{fig:badforcegoodimage}
    \end{subfigure}
    
    \caption{(a) Force signal, Ultrasound signal and labels in the presence of non-corrupted force and ultrasound data. (b) Force signal, Ultrasound signal and labels in the presence of noisy ultrasound data. (c) Force signal, Ultrasound signal and labels in presence of noisy force signal. }
    \label{fig:setup}
\end{figure*}

\subsection{Ultrasound Data Processing}
\label{sec:imageprocessing}
The informative component of the force signal (along y-axis ${F_y}$, Fig. \ref{fig:forcetrace}) is a 1D signal providing spatial information about the spine anatomy along the spine direction. However, ultrasound data are 3D data, where each position along the spine corresponds to a 2D (B-mode) ultrasound frame. To be able to effectively compare the information from the two sensors, ultrasound data are reduced to a 1D vector, defined along the spine direction. 
The dimension reduction is achieved by analyzing each ultrasound frame in the acquired sweeps and defining the probability for each of them to contain a vertebra.  The concatenation of the resulting values along the spine direction is a 1D signal where high probability peaks ideally coincide with vertebrae and therefore corresponds to peaks in the force signal.

The vertebra probability value is extracted from each frame using a Convolutional Neural Network trained for the task of classification. In order to ensure the best classification results, three state of the art classification networks were tested and compared: ResNet18 \cite{he2015deep}, DenseNet121 \cite{huang2016densely}, VGG11 with batch normalization \cite{Simonyan15}. The training and validation performances were evaluated for all the architectures in the following cases: 
a) Using ImageNet \cite{imagenet_cvpr09} weights as initialization (pre-trained network) and fine-tuning all layers.
b) Using ImageNet weights as initialization and fine-tuning the last layer only.
c) Training the network with randomly initialized weights.
Each model was trained using Adam optimizer, Cross-entropy loss function, learning rate of 0.0005 and a learning rate decay of 0.1 every 5 epochs for 30 epochs. 
The data for CNN training and testing were sampled from the Dataset 1. The training dataset consisted of 15 subjects (12 for training and 3 for validation), for a total of 1986 images for each class to ensure class balance. The test set consisted of 4 subjects, for a total of 696 images for each class. A 5-fold cross-validation study was performed over the training and validation datasets to exclude false-positive results. 
The obtained 1D signal is smoothed using a low-pass filtered with a second-order Butterworth with cutoff frequency at 0.3 Hz and re-sampled in equally spaced space-grid.

\subsection{Force - Ultrasound Data Fusion}
\label{sec:datafusion}
The extracted and pre-processed force and ultrasound 1D signals represent variations of the inner/outer spine anatomy along the spine direction.
In optimal conditions, both signals present well visible peaks in correspondence with vertebral levels (Fig. \ref{fig:goodforcegoodimg}). However, in some cases one (or both) signals may be corrupted by noise, making it challenging to identify the real position of the vertebral levels. Noise in the signal extracted from the ultrasound data typically arises from the scarce visibility of the spinous process in the ultrasound sweep (Fig. \ref{fig:goodforcebadimage}). This can be related to several factors as device-specific noise, non-optimal couplings between the probe and the patient skin or subject-specific anatomy and tissue distribution. Noise in the force signal may arise from sudden movements of the subject during the acquisition, or from subject-specific anatomical features (e.g. vertebral peaks may be less evident in particularly muscular subjects) (Fig. \ref{fig:badforcegoodimage}).

\begin{figure} []
    \centering
    \includegraphics[width=\linewidth]{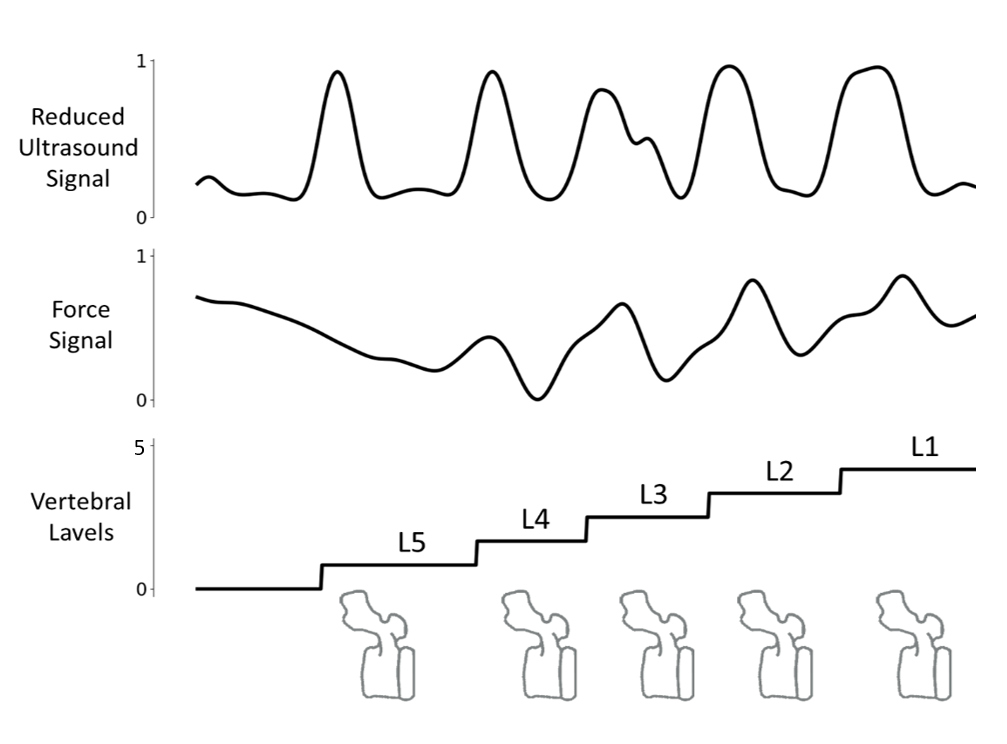}
    \caption{The first row represents the 1D probability signal extracted from the ultrasound data. The second row represents the force signal. The third row represents the corresponding, manually labelled vertebral levels and the step-wise function the data fusion method is trained on.}
    \label{fig:tcndata}
\end{figure}

To make the method more robust against single-sensor misclassifications, a force-ultrasound fusion method was implemented. 
In particular, a 1D Spatial Convolutional Network was trained to classify vertebral levels from the input signals.  The vertebral level counting problem is modelled as a classification problem, where the network is trained to classify each vertebral level in the lumbar region (Fig. \ref{fig:tcndata}).  

\begin{figure}
    \centering
    \includegraphics[width=\linewidth]{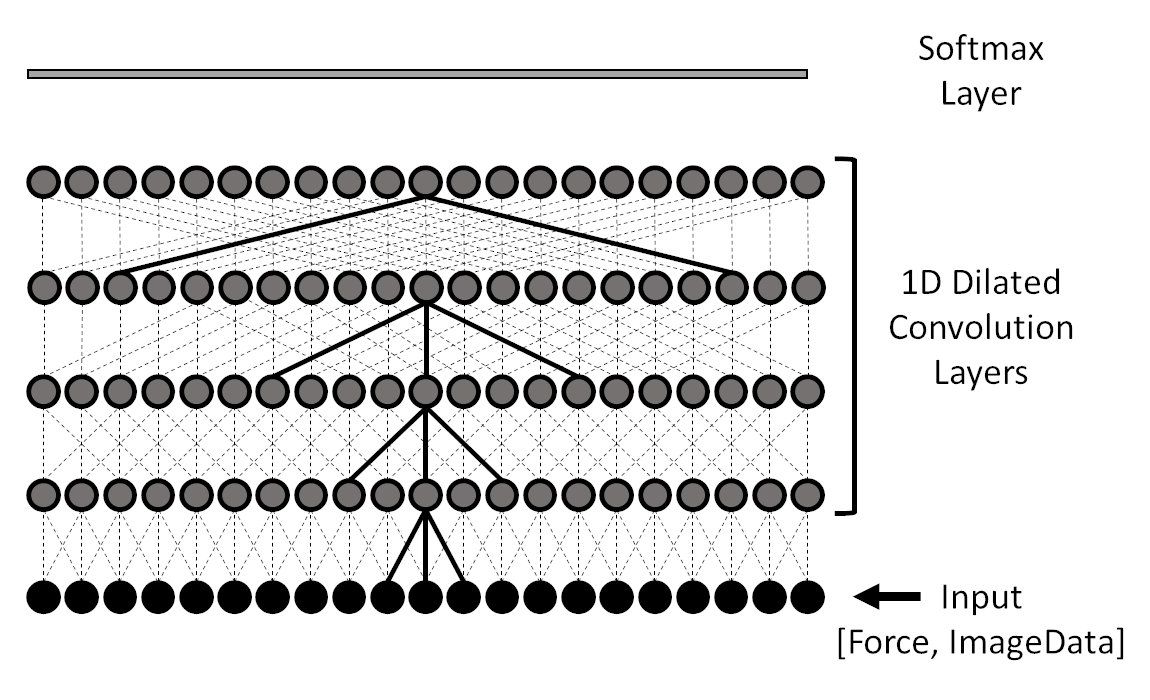}
    \caption{The architecture of the single stage of the 1D convolutional network for data fusion.}
    \label{fig:tcn}
\end{figure}

A multi-stage temporal convolutional network is devised based on \cite{farha2019ms}, where the overall architecture consists of three stages and each stage is trained to classify the input data. Each stage refines the results from previous stages, yielding smoother and more accurate classification results. Each stage consists of an initial 1x1 convolution layer which re-sizes the input into a 32 x N sequence, where N is the original signal length (number of samples along the spine direction). The initial layer is followed by 9 1xD dilated convolution layers with kernel size 3 and increasing dilation size (Fig. \ref{fig:tcn}). 
Dilated convolution is defined as: 
\begin{equation}
    (F *_l k)_t = \sum_{s+lt=p} F(S)k(t)
\end{equation}

where $F$ is the input signal, $k$ is the filter kernel and $l$ is the dilation factor. It can be seen from the formula that, compared to standard convolution,
the result at each point of the convoluted signal is obtained considering a larger spatial field in the input signal, therefore allowing the network to exploit a broader spatial context for the input's classification.
A softmax layer is added after the last convolution layer, to retrieve class probabilities (Fig. \ref{fig:tcn}). The cross-entropy and an additional smoothing factor are used as the loss function for network training, as described in \cite{farha2019ms}. 
The convolutional network for force and ultrasound fusion was trained using Adam optimizer, learning rate of 0.0005 and batch size 1 for 110 epochs. 
The data for network training and testing were sampled from Dataset 2. The training dataset consisted of multiple sweeps acquired over 9 subjects (7 for training and 2 for validation) sampled from Dataset 2, for a total of 27 sequences for training and 7 for validation. The test set consists of 4 unseen patients, acquired with the optimal robot parameters (force equal to 10N or 15N depending on subject BMI and robot speed equals to 20 mm/s). 

\section{Results}
\subsection{Evaluation}
\subsubsection{Ultrasound Data Processing}
\label{sec:results:CNN}
In Table, \ref{net_comparison} the test accuracy is reported for each CNN architecture (ResNet18, DenseNet121, VGG11) for the 3 training cases (using a pre-trained network with ImageNet weights as initialization and fine-tune all layers; training only the last layer of the network; training entire network with randomly initialized weights). 
The best accuracy on the test set is obtained by fine-tuning all the layers of ResNet18 from the pre-trained model, providing an average accuracy of ${0.929 \pm{0.006}}$. 
The ResNet18 model with the best performance was tested on a testing database of \numberOfTestSubjects subjects, yielding an overall accuracy of 0.938. The confusion matrix computed on the test data is displayed in Table \ref{tab:confusion matrix}. The values are normalized by the total number of frames, the number of images n = 1392, the correspondent number of frames is shown in the parenthesis.

{\renewcommand{\arraystretch}{1.5}
\begin{table}[]
\centering
\caption{\MakeUppercase{Results of 5-folds cross-validation study for various models with different training modes.}}

\begin{tabular}{ | C{2.3cm}| C{1.6cm} | C{1.6cm}| C{1.6cm} |}
\hline
\textbf{} & \textbf{ResNet18} & \textbf{DenseNet121} & \textbf{VGG11-bn}\\
\hline
Training with randomly initialized weights  &  \textbf{$0.817 \pm{0.118}$}  & \textbf{$0.878 \pm{0.047}$} & \textbf{$0.635 \pm{0.15}$}\\
\hline
Pre-trained weights \& all layers fine-tuned &  \textbf{$0.929 \pm{0.006}$} & \textbf{$0.89 \pm{0.014}$}& \textbf{$0.878 \pm{0.055}$}\\
\hline
Pre-trained weights \& last layer only fine-tuned&  \textbf{$0.6 \pm{0.02}$} & \textbf{$0.577 \pm{0.006}$}& \textbf{$0.63 \pm{0.03}$} \\
\hline
\end{tabular}

\label{net_comparison}
\end{table}
}

\begin{table}[]
\centering
\setlength{\belowcaptionskip}{10pt}
\caption{\MakeUppercase{Confusion matrix for the best model performance evaluated on the test set of 4 subjects.}}

\begin{tabular}{@{}ccccl@{}}
                        &          & \multicolumn{2}{c}{Predicted}                                                                                                                                          &  \\
                        & n = 1392 & Vertebra                                                                                     & Intervertebral Gap                                                                     &  \\ \cmidrule(lr){2-4}
\multirow{2}{*}{Actual} & Vetrebra & \multicolumn{1}{c|}{\begin{tabular}[c]{@{}c@{}}True Positive\\ 0.459 (n = 640)\end{tabular}} & \begin{tabular}[c]{@{}c@{}}False Negative\\ 0.04 (n = 30)\end{tabular} &  \\ \cmidrule(lr){2-4}
                        & Intervertebral Gap      & \multicolumn{1}{c|}{\begin{tabular}[c]{@{}c@{}}False Positive\\ 0.02 (n = 30)\end{tabular}} & \begin{tabular}[c]{@{}c@{}}True Negative\\ 0.478 (n = 666)\end{tabular} &  \\ \cmidrule(lr){2-4}
                        & &  &  & 
\end{tabular}
\label{tab:confusion matrix}

\end{table}

\subsubsection{Force-Ultrasound Data Fusion}
The performances of the force-ultrasound data fusion method were evaluated in terms of its capability to correctly label each vertebral level. The test group consists of 5 (unseen) subjects, for a total of 20 vertebral levels.
To proof the robustness of the method and the effectiveness of combining different sensor data for vertebral level counting, the force-ultrasound fusion method was compared against pure ultrasound-based and pure force-based approaches. Pure force and pure image-based methods were obtained using the same network architecture described in Sec. \ref{sec:datafusion}, using the force and image signals alone as input data.
In order to simulate a realistic environment in the test phase, the offline data were streamed to the main workstation with proper streaming frequencies (30 fps for both ultrasound and force sensor).

In Table \ref{tab:countingresults} the results of the vertebral classification are reported for the three methods, as well as the distance from the ground truth vertebral level. A vertebral level classification is considered successful if an overlap higher than 0.5 exists between labels and predictions. It can be seen that the fusion method outperforms both pure force and pure image-method in the task of classification. 

\begin{figure}[]

\includegraphics[width=\linewidth]{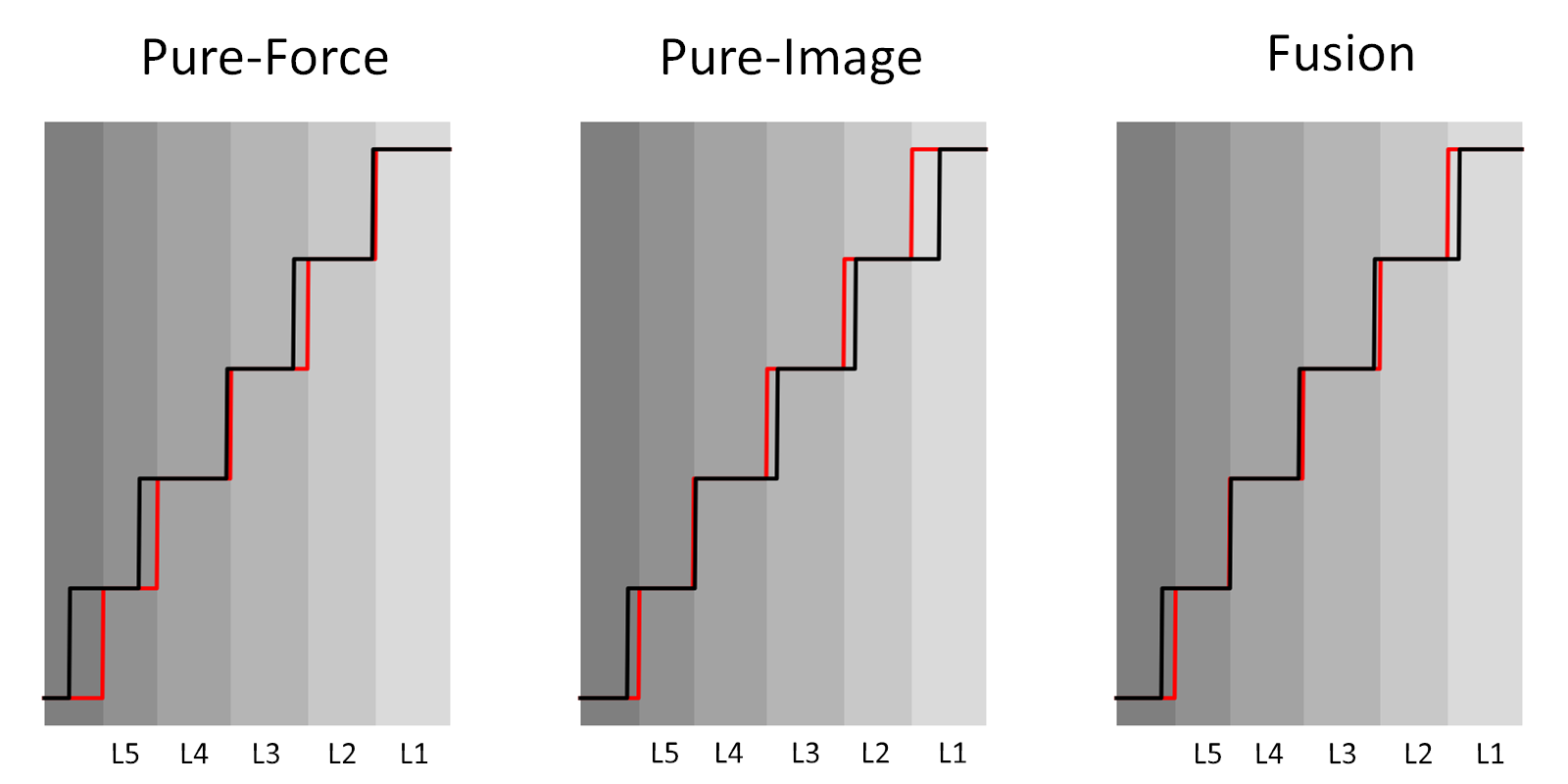}
  \caption{The predicted (black line) and ground-truth (red line) vertebral levels for pure force-based, pure ultrasound-based and force-ultrasound fusion method in an optimal case where both ultrasound and force signals are not noisy or corrupted (Subject Gender: Male, BMI: 25).}
  \label{fig:results1}
\end{figure}

In Fig. \ref{fig:results1} the results for the three methods are shown for the optimal case where both the force and ultrasound signals are not corrupted by noise. It can be seen that in this case, using force or ultrasound data alone is sufficient for obtaining a precise classification and counting of the vertebral levels.

\begin{table}[]
\centering
\caption{\MakeUppercase{The classification performances and distance from the ground truth vertebrae position for the three tested methods.}}
\begin{tabular}{ | C{2.3cm}| C{1.5cm} | C{1.5cm}| C{1.5cm}|}
\hline
 & Pure Force &Pure Ultrasound &Fusion \\
\hline
Correctly Classified Levels &  18/20 & 16/20 & 20/20 \\
\hline
Distance from Ground Truth Label [mm] &5.97 $\pm$ 5.91 &3.22 $\pm$ 3.07 & 3.47 $\pm$ {2.78} \\
\hline
\end{tabular}

\label{tab:countingresults}
\end{table}

\begin{figure}[]

    \includegraphics[width=\linewidth]{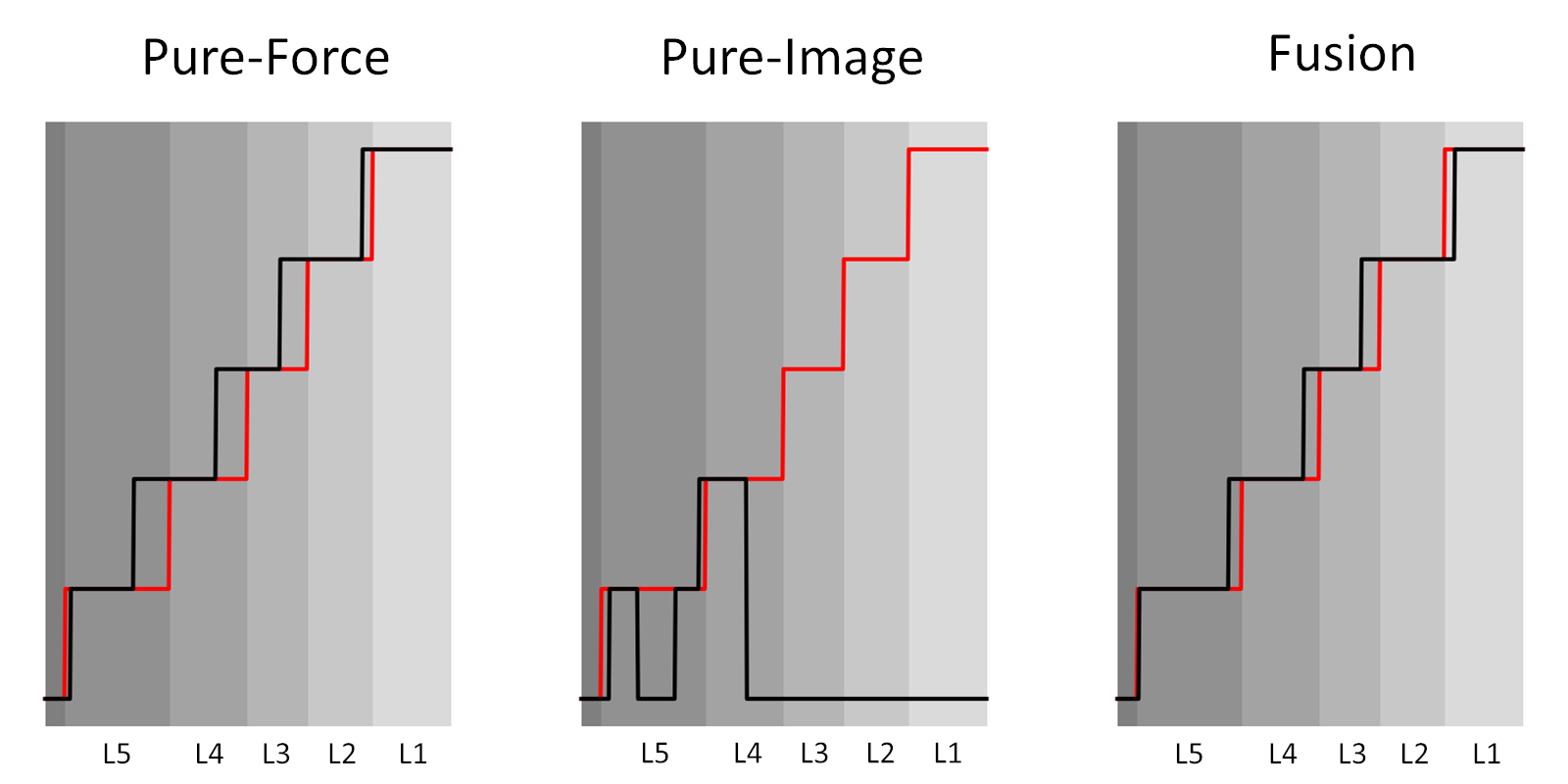}
  \caption{The predicted (black line) and ground-truth (red line) vertebral levels for pure force-based, pure ultrasound-based and force-ultrasound fusion method in a case where the ultrasound signal is noisy due to scarce visibility of the spinous process (Subject Gender: Male, BMI: 30).}
  \label{fig:results2}
\end{figure}

 In Fig. \ref{fig:results2} the results for the three methods are shown in the presence of noisy ultrasound data. It can be seen that the pure force-based method is able to correctly classify the vertebral levels. However, the pure ultrasound-based method fails to classify the last 3 vertebral levels. The fusion between the two methods is able to compensate for the ultrasound signal misclassification and to correctly classify all the lumbar vertebral levels. 
 
\begin{figure}[]

    \includegraphics[width=\linewidth]{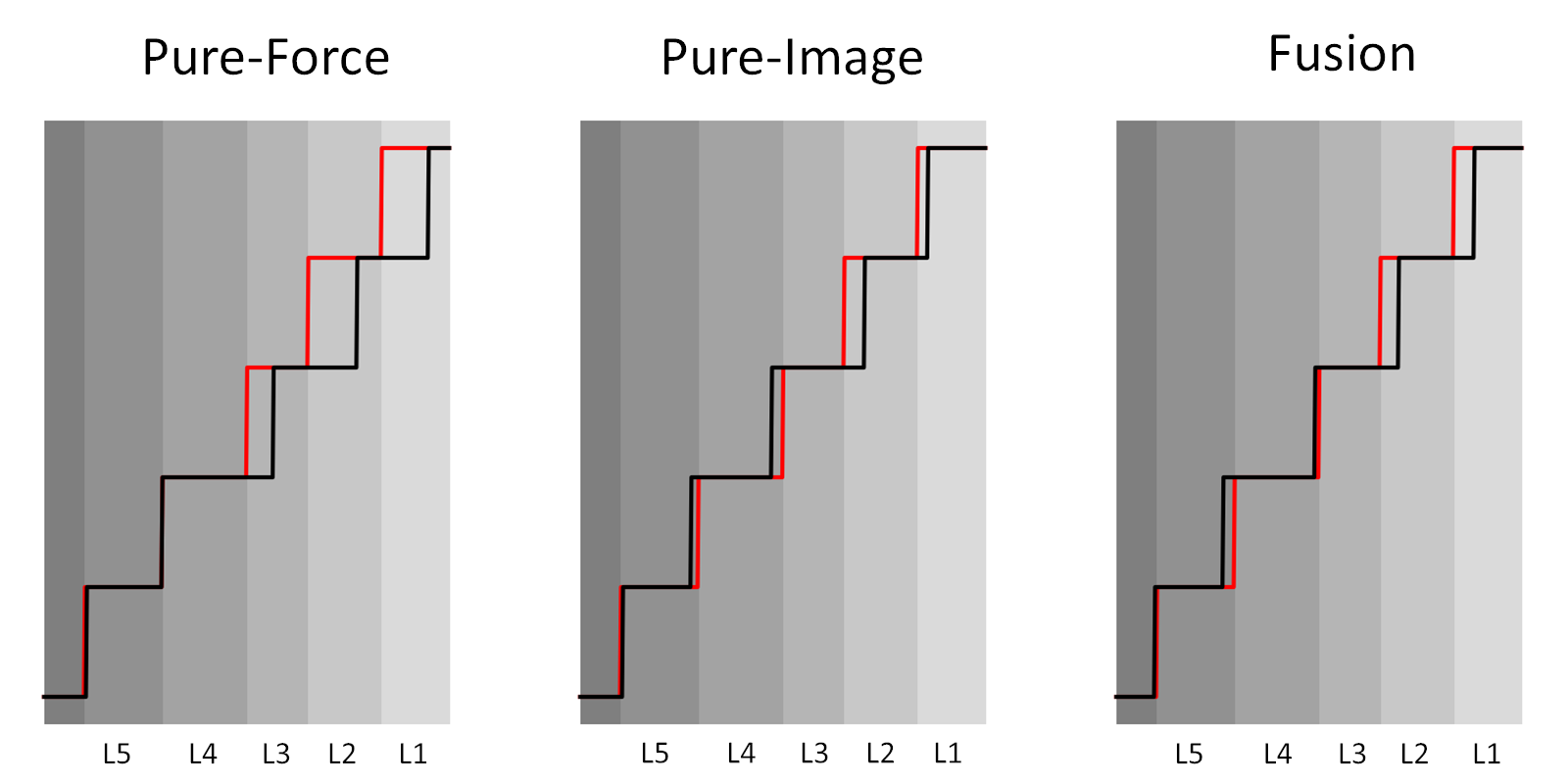}
  \caption{The predicted (black line) and ground-truth (red line) vertebral levels for pure force-based, pure ultrasound-based and force-ultrasound fusion method in a case where the force signal is noisy (Subject Gender: Female, BMI: 20). }
  \label{fig:results3}
\end{figure}

In Fig. \ref{fig:results3} the results for the three methods are shown in the presence of noisy force data. It can be seen that the pure force-based method fails in the classification of the last two vertebral levels while the pure ultrasound-based method is able to correctly classify them. Even in this case, the fusion method is able to correctly classify the input data, even in the presence of force signal corruption.

\subsubsection{The potential application}

The performances of the presented method were tested for an example application, i.e. automatic target plane selection for facet injection procedures. The facet injection procedure is performed to deliver anaesthetics at the level of facet joints, i.e. the anatomical structures connecting consecutive vertebrae (Fig. \ref{fig:scanningprocedure_subplot_facet}).
Using the proposed vertebral level classification method, the correct vertebral level can be selected, and a sweep can be taken at the correct level with the probe in a paramedian sagittal orientation, to identify the target injection plane.

The method for facet plane identification is similar to the one presented in \cite{Pesteie2015}. 
Each frame in the sweep is classified as either ``facet" or ``non-facet" plane and the two frames with the highest probability in the sweep are labelled as right and left facet planes. The plane classification task is performed using ResNet18, given its high performances in the ultrasound classification task (Sec. \ref{sec:results:CNN}).

The model was pre-trained on ImageNet and fine-tuned on a training set sampled from Dataset 3. 
The spatial errors between identified facet joint planes and labelled planes were calculated on 4 test subjects sampled from Dataset 3, which consisted of 20 vertebrae sweeps (5 vertebrae for each subject), each containing two facet joints, resulting in 40 facet joints in total. For 37 facet joints out of 40, the mean distance error between the detected and manually labelled facet planes is $2.08 \pm{2.63}$ mm. According to \cite{Greher2004} an error below 5 mm leads to an effective anaesthetic result for the facet joint injections. For the rest 3 facet joints out of 40, the error is $8.43 \pm{8.98}$ mm since the CNN output resulted to be less precise, due to the poor image quality.

\section{Conclusion}

Currently, clinical routine spine injections procedures
completely rely on the expertise of the surgeon, both to ensure the accuracy of the procedure and to limit the exposure time  to the ionizing radiation. In this study, a robotic-ultrasound method for vertebral level detection and counting was developed for spine injection procedures. To the best of our knowledge, it is the first robotic system integrating visual and force feedback for vertebra level classification.

The method was tested on a group of healthy volunteers, chosen to maximize the inter-subject variability in terms of gender and BMI. However, a more thorough analysis should be conducted with a larger database, to better understand the correlation between method performances and subject anatomical characteristics. Future exploration should also focus on the online validation of the method in a real clinical environment and further automation of each step of the injection procedure. Furthermore, possible application in other clinical scenario as scoliosis assessment should be considered, since accessing the position of each vertebral level is beneficial for curvature reconstruction \cite{Victorova2019}.

The proposed method effectively fuses ultrasound and force data acquired during a robotic actuated scanning of the patient back for vertebral level classification. It was proven that the proposed fusion method yielded higher performances compared to pure image and pure force-based methods. From the results, it can be noticed that, when able to classify a vertebral level, the pure image method can precisely detect its correct location. However, in cases where the input ultrasound image is particularly noisy, it totally fails to detect vertebrae.
Even when using the pure force-based method, corruption in the input force data may lead to misclassification of the vertebral levels. By combining image and force data, the fusion method is able to correctly classify vertebral level even in presence of force or ultrasound data corruption, with a precision comparable to the one obtained with the pure image method. In particular, the fusion method correctly classifies 100\% of the vertebral levels in the test set with a precision of 3.47 mm, while pure image and pure force-based method could only classify 16 and 18 out of 20 vertebrae with a precision of 3.22 mm and 5.97 mm, respectively.

The potential of the proposed method was explored in the integration with a common clinical procedure, opening the path for future exploration toward fully automatic injection procedures.


\bibliographystyle{IEEEtran}
\bibliography{IEEEabrv,root.bib}

\end{document}